\documentclass[preprint,amsmath,amssymb,letterpaper,onecolumn]{revtex4}
\pdfoutput=1

\usepackage{amsmath}
\usepackage{graphicx}
\usepackage[margin=2cm]{geometry}
\def\beq{\begin{equation}}
\def\eeq{\end{equation}}

\def\kt{k_{\rm B}T}

\def\be{\begin{equation}}
\def\ee{\end{equation}}
\def\ba{\begin{eqnarray}}
\def\ea{\end{eqnarray}}

\begin{document}

\linespread{1.6}
\title{Microscopic implications of S-DNA}
\author{Stephen Whitelam\footnote{\texttt{swhitelam@lbl.gov}}$^1$, Phillip L. Geissler$^2$ and Sander Pronk$^3$}
\affiliation{$^1$The Molecular Foundry, Lawrence Berkeley National Laboratory, Berkeley, CA 94720, USA\\
$^2$Department of Chemistry, University of California at Berkeley, Berkeley, CA 94720, USA\\
$^3$Center for Biomembrane Research, Department of Biochemistry \& Biophysics, Stockholm University, SE-106 91 Stockholm, Sweden}
\begin{abstract}
\linespread{1.2}
Recent experiments [J. van Mameren {\em et al.} PNAS 106, 18231 (2009)] provide a detailed spatial picture of overstretched DNA, showing that under certain conditions the two strands of the double helix separate at about 65 pN. It was proposed that this observation rules out the existence of an elongated, hybridized form of DNA (ÔS-DNAÕ). Here we argue that the S-DNA picture {\em is} consistent with the observation of unpeeling during overstretching. We demonstrate that assuming the existence of S-DNA does not imply DNA overstretching to consist of the complete or near-complete conversion of the molecule from B- to S-form. Instead, this assumption implies in general a more complex dynamic coexistence of hybridized and unhybridized forms of DNA. We argue that such coexistence can rationalize several recent experimental observations.
\end{abstract}
\maketitle

\linespread{1.2}

\section{Introduction}

\noindent
Torsionally unconstrained, double-stranded DNA under longitudinal tension undergoes an abrupt elongation or `overstretching' at about 65 pN, attributed either to melting~\cite{bloomfield1,bloomfield0,bloomfield2} or to the emergence of an elongated, hybridized state called S-DNA~\cite{bustamante,s_form1,cocco,s_form3,s_form2}. Recent elegant experiments~\cite{van2009unraveling} show directly that stretched, torsionally unconstrained molecules strand-separate or `unpeel' near 65 pN, an observation argued to rule out the existence of S-DNA. Here we argue instead that the S-DNA picture {\em is} consistent with the observation of unpeeling during stretching. Specifically, we demonstrate that this picture predicts that at low temperatures, high salt concentrations and when large forces are applied only for short times, S-DNA may be observed, but that under the converse conditions -- those probed in Ref.~\cite{van2009unraveling} -- unpeeling will coexist with or occur in preference to S-DNA. We argue further that dynamic coexistence of S-DNA and unpeeled DNA can rationalize several recent experimental observations.\\

\noindent
The essence of the S-DNA picture of overstretching~\cite{bustamante,s_form1,cocco,s_form3,s_form2} is the conjecture that there exists an elongated, hybridized phase of DNA distinct from B-DNA. While cartoon sketches of this picture may convey the idea that it predicts overstretching to consist of the complete conversion of the molecule from B- to S-form, thermodynamic and kinetic analysis~\cite{cocco,us, whitelam2008stretching} predicts a more complex scenario of dynamic coexistence between hybridized and unhybridized forms of DNA. The nature of this coexistence is strongly dependent upon conditions of salt and temperature -- low salt concentrations and high temperatures favoring unpeeling -- and upon stretching dynamics. \\

\noindent
Here we use the coarse-grained, statistical mechanical model of Refs.~\cite{us,whitelam2008stretching} to make predictions for the molecular composition of a 40 kbp fragment of nicked $\lambda$-DNA under tension, for a range of environmental conditions, assuming the existence of S-DNA. We find that the thermodynamically-favored composition of DNA under tension (summarized in Fig.~\ref{fig1}) is very different from the composition that results if the molecule is stretched dynamically (Figs.~\ref{fig3} to~\ref{fig6}). This difference results from the very slow emergence of unpeeled DNA~\cite{cocco}. We find that the instantaneous fraction of S-form DNA varies strongly with stretching force, ambient temperature and salt concentration, and observation time. At the 50 mM salt concentration and (presumably, room) temperatures considered in Ref.~\cite{van2009unraveling} we find S-DNA to be very short-lived, and likely unobservable in stretching experiments lasting a few seconds or more. At low temperatures and/or high salt concentrations, S-DNA can be long-lived. These predictions are in accord with those of Ref.~\cite{cocco}, and with recent experimental evidence for the existence of S-DNA~\cite{fu2010two}. The latter work identifies two DNA overstretching `modes': a fast mode that operates principally at low temperatures, high salt concentrations and in CG-rich sequences, and a slow mode that proliferates under the converse conditions. These observations were argued to be in accord with the predictions of the S-DNA picture advanced in Ref.~\cite{cocco}, with the fast mode reflecting the emergence of S-DNA and the slow mode signaling unpeeling.

\section{Model and simulation details}

\noindent
The model we consider resolves a given basepair (of type A, T, C or G) only to the extent that we assign to it a particular instantaneous state. We consider four possible states, sketched in Fig.~\ref{fig1}(a). The first is the helical B-form of DNA. The second is the `molten bubble' form of DNA (M-DNA) in which both strands bear tension but do not interact with each other. The third state is the unpeeled form (U-DNA) in which one strand has frayed and no longer bears tension. It is crucial to distinguish between the two forms of unhybridized DNA: in the presence of nicks, M-DNA is unstable to U-DNA~\cite{cocco}. All three states can be parameterized using thermal and mechanical data~\cite{bustamante,cocco, santa}. The fourth state is the putative S-form (S-DNA). The model of S-DNA used in Refs.~\cite{us, whitelam2008stretching} is based on two assumptions: first, that S-DNA has a contour length 70\% greater than B-DNA and is stiffer at large forces~\cite{storm2003theory,cocco,calderon2009quantifying}; and second, that it is in thermal equilibrium with B-DNA, irrespective of sequence composition, at 65 pN. The motivation for the latter assumption is the observation that $\lambda$-DNA (about 50\% CG content) and poly(dG-dC) DNA both overstretch at about 65 pN~\cite{cs}. This observation is difficult to reconcile with the assumption that overstretching consists solely of melting of the double helix. \\

\noindent
The thermodynamics implied by this picture at a salt concentration of 150 mM is summarized in Fig.~\ref{fig1}(b) (see Appendix). At low forces B-DNA is stable; at high forces either U-DNA or S-DNA are stable. The assumption that S-B phase coexistence occurs at 65 pN independently of sequence and environmental conditions immediately implies that S-DNA is more stable with respect to U-DNA in CG-rich DNA than in AT-rich DNA, and that increasing temperature favors U-DNA over S-DNA. Further, for the parameters used in Ref.~\cite{us}, we find that S-DNA is not thermodynamically stable (under the conditions shown) with respect to U-DNA within the first 40 kbp of $\lambda$-DNA as a whole. However, as discussed in Ref.~\cite{cocco}, this does not mean that S-DNA is never observed in dynamic stretching simulations. U-DNA must emerge in a processive, basepair-by-basepair fashion. It is therefore possible to find thermodynamic conditions such that while complete unpeeling of the molecule is thermodynamically favorable, the passage of U-DNA through CG-rich tracts of the molecule is hindered by free energy barriers in excess of hundreds of $ k_{\rm B} T$. Under such conditions, S-DNA can be transiently stable. \\

\noindent
To illustrate this point, we show in Fig.~\ref{fig2} a microscopic space-versus-time plot of dynamic molecular composition within a 300-basepair fragment of $\lambda$-DNA stretched at 1000 nm/s. Unpeeling from the molecule's free end eventually leads to complete dehybridization, but because the advance of the unpeeled front is processive and slow, S-DNA is transiently stable. To reveal the effect of environmental conditions on the degree of this stability, we plot in Fig.~\ref{fig3} a `dynamic phase diagram' at salt concentration $c=150$ mM for 40 kbp fragments of $\lambda$-DNA stretched at 1000 nm/s at temperature $T$ to a target force $f_0$. We use the `optical trap' protocol described in detail in Ref.~\cite{us}: trap stiffness is 1 pN/nm, and molecules possess one free end and no internal nicks. Once the target force is achieved, the loading rate is set to zero and the molecular composition monitored. We show molecular composition (vertical axis, maximum height represents 40 kbp) at a range of thermodynamic states at times 0 and 10 seconds after setting the loading rate to zero. Two features are apparent. First, S-DNA is sufficiently long-lived to be seen after 10 seconds of waiting at low temperatures. Second, near and above room temperature, S-DNA is short-lived. We show a similar diagram in Fig.~\ref{fig7} for identical thermodynamic conditions but a smaller stretching rate of 100 nm/s. Because the target force $f_0$ is attained more slowly, and molecules spend more time at high forces than do their counterparts in Fig.~\ref{fig3}, more unpeeling is observed. We show additional `dynamic phase diagrams' at salt concentrations of 50 mM (Fig.~\ref{fig4}) and 500 mM (Fig.~\ref{fig5}), and individual molecular compositions as a function of sequence in Fig.~\ref{fig6}. We find that S-DNA is short-lived and that unpeeling occurs readily under the conditions considered in Ref.~\cite{van2009unraveling}: we would not expect S-DNA to be observed at forces near 65 pN after a few seconds under those conditions. Repeating such experiments at high salt concentrations and at low temperatures would provide a test of the predictions presented here.\\
 
\noindent
We have argued previously~\cite{us,whitelam2008stretching} that the transient emergence of S-DNA rationalizes several recent experimental observations. The basepair-by-basepair processivity of U-DNA means that its emergence and disappearance is slow, which renders hysteretic a stretching-shortening cycle~\cite{bustamante,cocco}. Moreover, competition between S-DNA's basepairing-stacking energy and U-DNA's entropy confers a strong temperature dependence upon the degree of hysteresis accompanying a stretching-shortening cycle, in agreement with the striking kinetic data of Ref.~\cite{hanbin}. The transient stability of S-DNA also rationalizes~\cite{cocco,us} a multi-stage force-extension behavior (for DNA at sufficiently low temperature, high salt concentration and possessing sufficiently low AT content), in which a pulling rate-independent plateau at 65 pN is succeeded at high forces by a second, rate-dependent transition, as seen experimentally~\cite{rief,cs}. By contrast, we predict that DNA forced to dehybridize during overstretching (as would happen if S-DNA does not exist) gives rise to considerable hysteresis under {\em all} solution conditions (rather than only at high temperatures) and a strongly pulling rate-dependent overstretching force~\cite{us}. Finally, we note that the `force-induced melting' picture proposes that overstretched DNA near 65 pN consists of internal molten bubbles (M-DNA) punctuated by small regions of B-DNA, and that unpeeling occurs at higher forces~\cite{bloomfield1,bloomfield0,bloomfield2}. The observation of Ref.~\cite{van2009unraveling} that DNA at low salt concentrations progressively {\em unpeels} near 65 pN from nicks and free ends appears to challenge this picture. 

\noindent
\section{Summary}

\noindent
Statistical mechanical considerations cannot suggest likely atomistic structures for S-DNA (possible manifestations of which have been reported in recent atomistic simulations~\cite{luan2008strain,li-overstretching}). However, we argue that overstretching dynamics revealed by experiments, especially the recent Ref.~\cite{fu2010two}, implies the existence of an elongated phase of DNA {\em energetically} preferred to U-DNA. We have noted previously the limitations of the coarse-grained picture that we advocate, as well as the fact that a different interpretation of M-DNA (does it retain some energetic advantage over U-DNA, for instance?) might yield agreement between our model calculations and experimental data in the absence of the putative S-phase. However, we stress that if experiments are to conclusively rule for or against either picture of overstretching, then the consequences of both must be better understood. This paper is an attempt to clarify the physical consequences of assuming the existence of an elongated, hybridized phase of DNA.

\noindent
\section{Acknowledgements} 

\noindent
We thank Gijs Wuite, Erwin Peterman and John Marko for correspondence. Fig.~\ref{fig2} was adapted from Ref.~\cite{whitelam2008stretching}. This work was supported by the Director, Office of Science, Office of Basic Energy Sciences, of the U.S. Department of Energy under Contract No. DE-AC02--05CH11231.

\clearpage

\noindent
\section{Appendix}

\begin{figure}
\label{}
\centering
\includegraphics[width=0.85\linewidth]{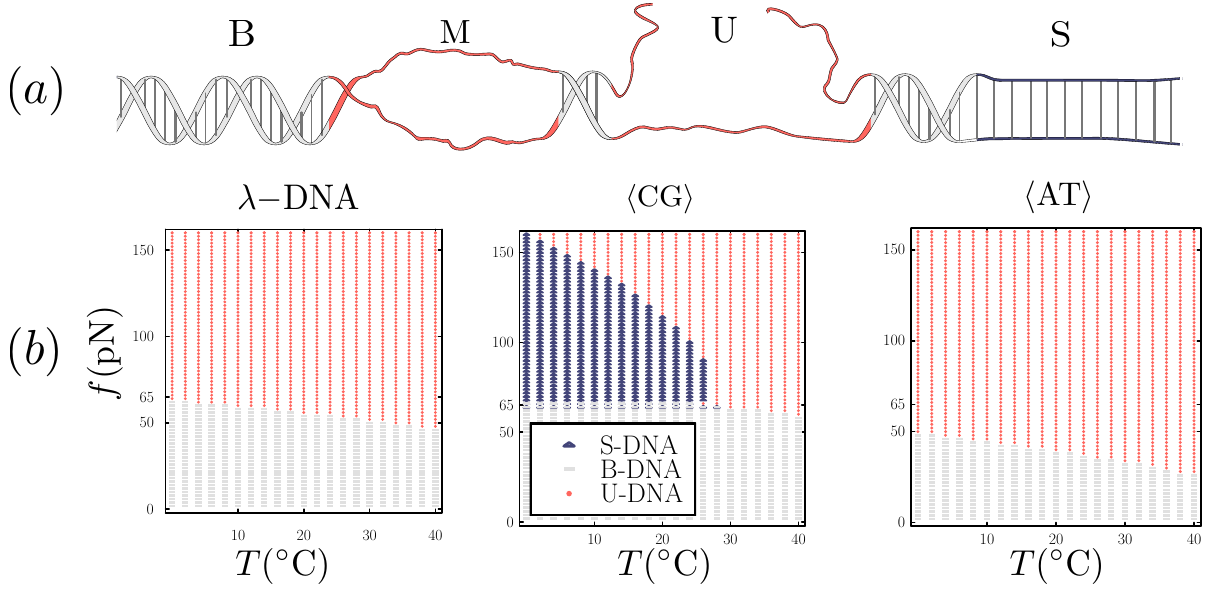} 
\caption{\label{fig1} (Color) The thermodynamics implied by the S-DNA picture of overstretching for salt concentration $c=150$ mM. (a) Cartoon sketches of phases B, M, U and S (see text). (b) Equilibrium phase stabilities (see Appendix) in the temperature-pulling force plane for phases U, B and S. We consider a 40 kbp fragment of $\lambda$-DNA (left), a similar length of `averaged' CG-DNA (center; see Appendix for details of sequence averaging), and averaged AT-DNA (right). The model of S-DNA described in the main text is not thermodynamically stable at any temperature or stretching force within the first 40 kbps of $\lambda$-DNA as a whole. However, its considerable stability against unpeeling in CG-rich regions suggests the possibility of long-lived metastable `pockets' of S-DNA during dynamic stretching experiments if nicks are few. Compare these data with the `dynamic phase diagram' of Fig.~\ref{fig3}.}
\end{figure}

\begin{figure}
\label{}
\centering
\includegraphics[width=0.85\linewidth]{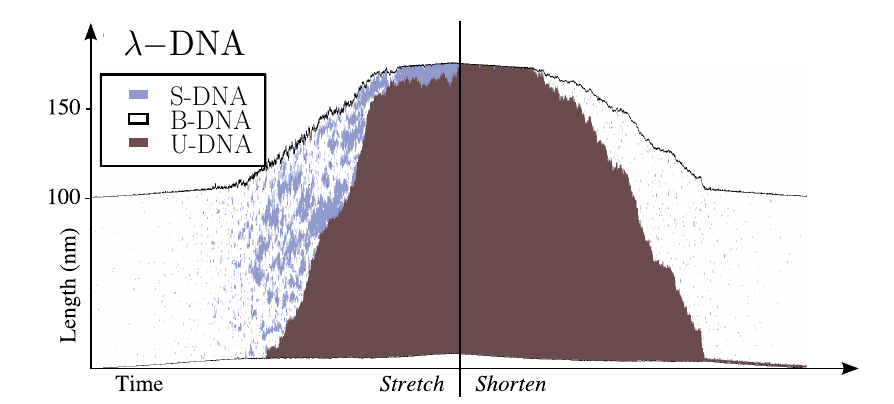} 
\caption{\label{fig2} (Color) Dynamic simulations of a microscopic model accommodating U, M, B and S forms of DNA~\cite{whitelam2008stretching}. Here we show microscopic configurations (vertical axis) as a function of time (horizontal axis) from an illustrative simulation of a 300-basepair fragment of $\lambda$-DNA at $21^{\circ}$C and salt concentration 150 mM, extended at 1000 nm/s. The molecule possesses one free end, permitting unpeeling. White indicates B-DNA, red indicates unhybridized DNA (unpeeled or molten), and blue indicates S-DNA. The spatial scale indicates molecular length. The maximum force attained is 90 pN (central dotted line). At forces in excess of about 50 pN, S-form DNA is themodynamically unstable with respect to unpeeled DNA. However, unpeeling takes time to proliferate, and S-DNA is seen to be transiently stable even near forces of about 90 pN. The step-by-step processivity of unpeeling confers upon overstretching an anomalous kinetics that, we argue, rationalizes several nontrivial experimental observations~\cite{cs,rief,hanbin}.}
\end{figure}

\begin{figure}
\label{}
\centering
\includegraphics[width=0.9\linewidth]{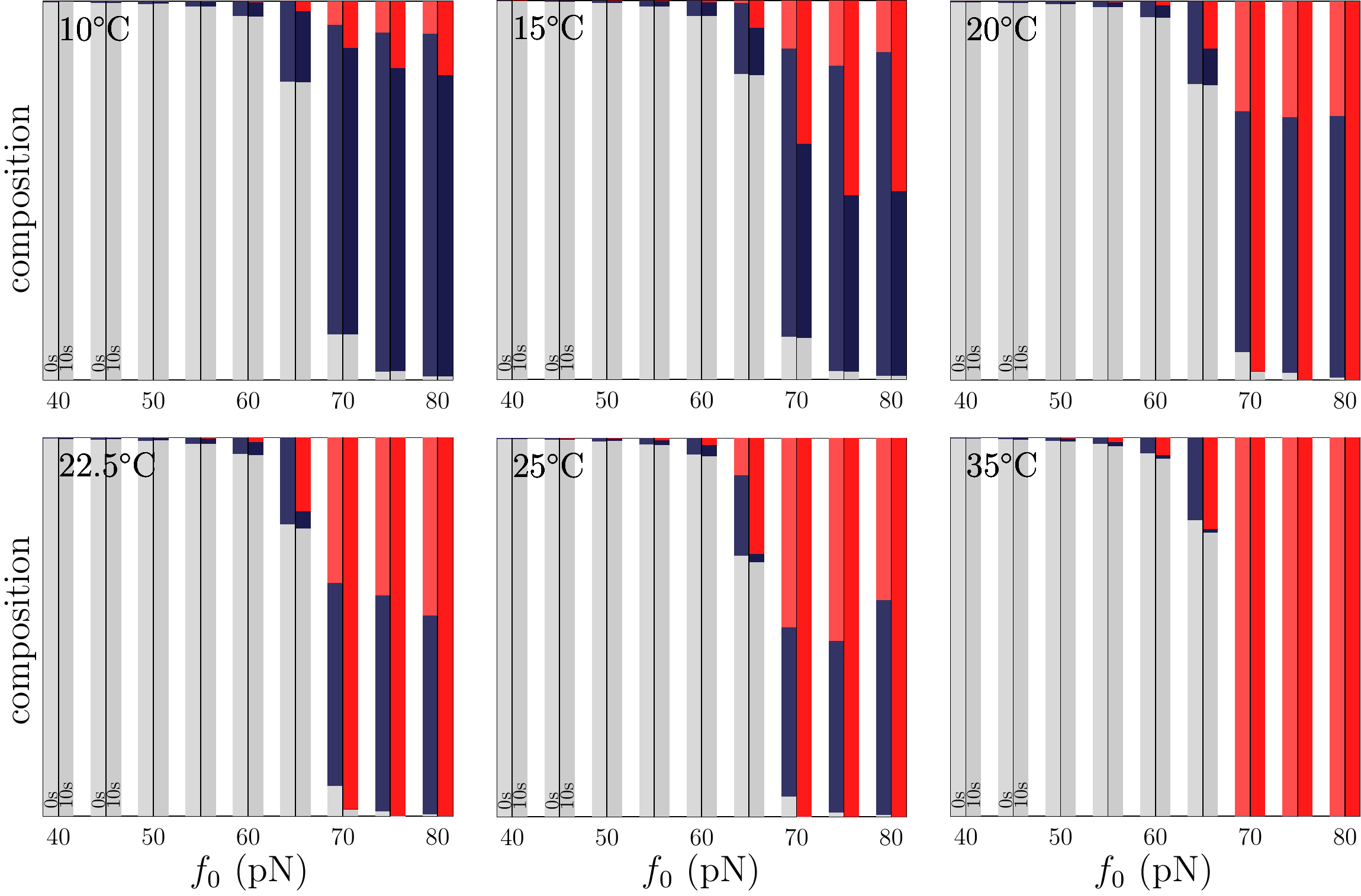} 
\caption{\label{fig3} (Color) `Dynamic phase diagram' for 40 kbp fragments of $\lambda$-DNA with one free end, at salt concentration $c=$150 mM: compare the thermodynamic phase diagram of Fig~\ref{fig1}. Color code: B-DNA (off-white), U-DNA (red), and S-DNA (blue). Here we use the `optical trap' protocol described in Ref.~\cite{us} (trap stiffness 1 pN/nm) to stretch DNA at fixed temperature $T$ at a rate of 1000 nm/s until a force $f_0$ is achieved. The loading rate is then set to zero, and the composition of the molecule is monitored (both molecular force and extension may vary, and force typically decreases slowly). We show molecular composition (fraction of B-DNA (off-white), U-DNA (red), and S-DNA (blue) denoted by bars on the vertical axis, where maximum height represents 40 kbp) for a molecule at specified temperature $T$ for target force $f_0$ at times 0 and 10 seconds after setting the loading rate to zero. Two features are apparent. First, S-DNA is sufficiently long-lived to be seen after 10 seconds of waiting at low temperatures. Second, near and above room temperature, S-DNA is short-lived.}
\end{figure}

\begin{figure}
\label{}
\centering
\includegraphics[width=0.9\linewidth]{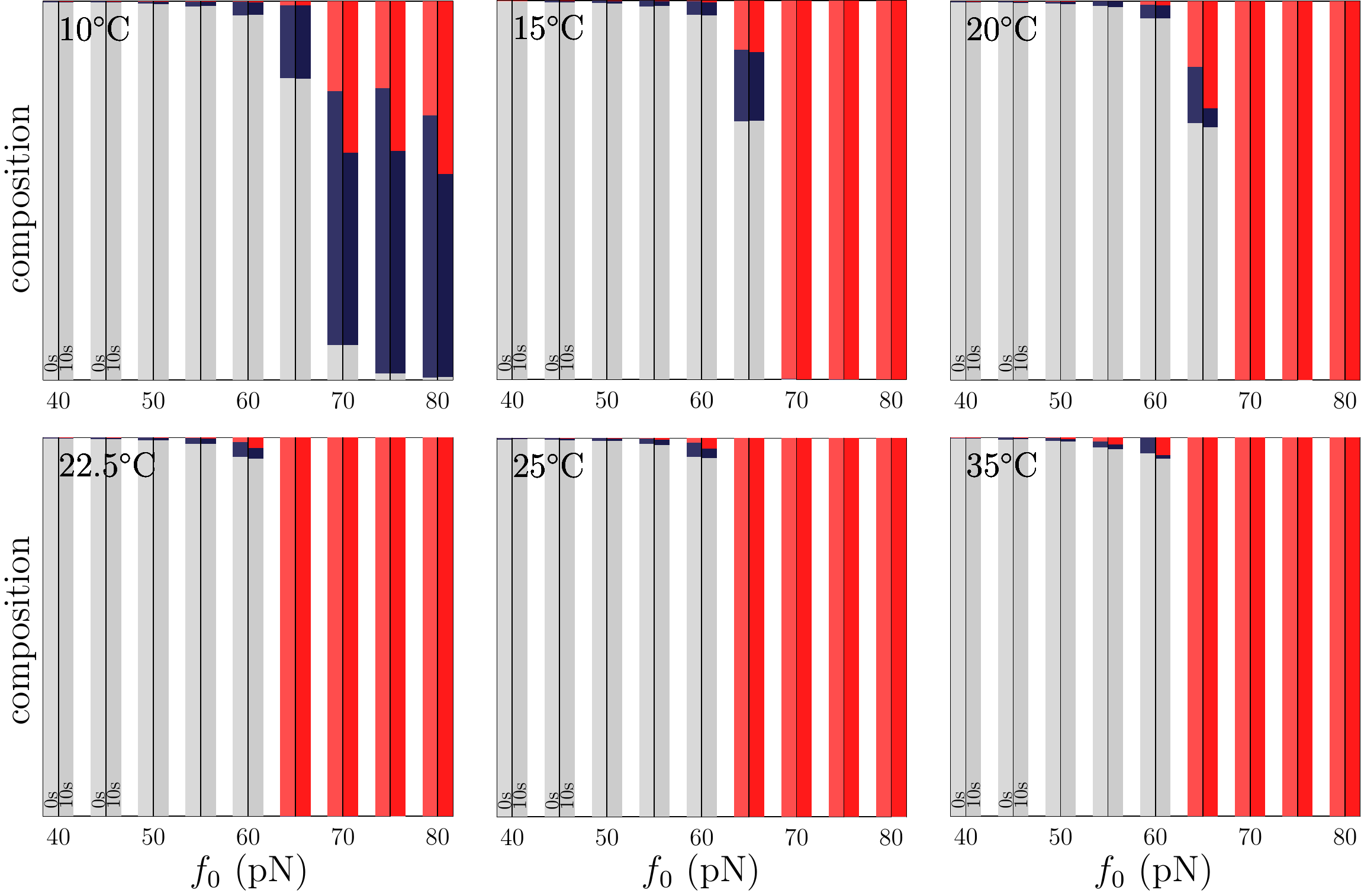} 
\caption{\label{fig7} (Color) `Dynamic phase diagram' for 40 kbp fragments of $\lambda$-DNA with one free end, at salt concentration $c=$150 mM. The dynamic stretching protocol is as Fig.~\ref{fig3}, except that the molecular stretching rate is 100 nm/s rather than 1000 nm/s. In general, more U-DNA is seen here then in Fig.~\ref{fig3}: here, attainment of the target force $f_0$ takes of order hundreds of seconds near 65 pN, and as a result the molecule has more time to unpeel than it does at the smaller stretching rate.}
\end{figure}

\begin{figure}
\label{}
\centering
\includegraphics[width=0.9\linewidth]{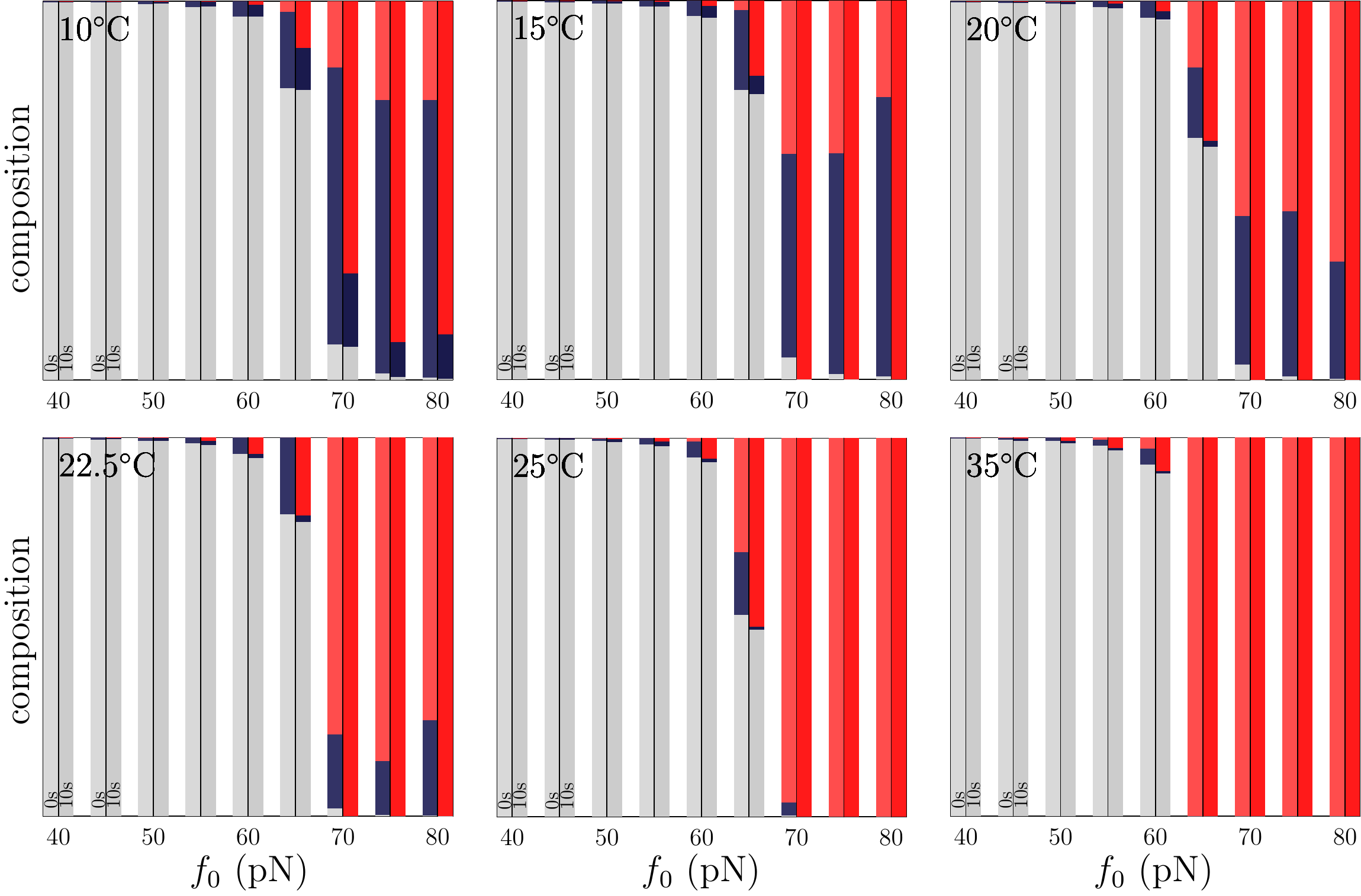} 
\caption{\label{fig4} (Color) `Dynamic phase diagram' for 40 kbp fragments of $\lambda$-DNA with one free end, at salt concentration $c=$50 mM. The dynamic stretching protocol is as Fig.~\ref{fig3}: note the smaller fraction of S-DNA seen here. At room temperature, S-DNA is short-lived.}
\end{figure}

\begin{figure}
\label{}
\centering
\includegraphics[width=0.9\linewidth]{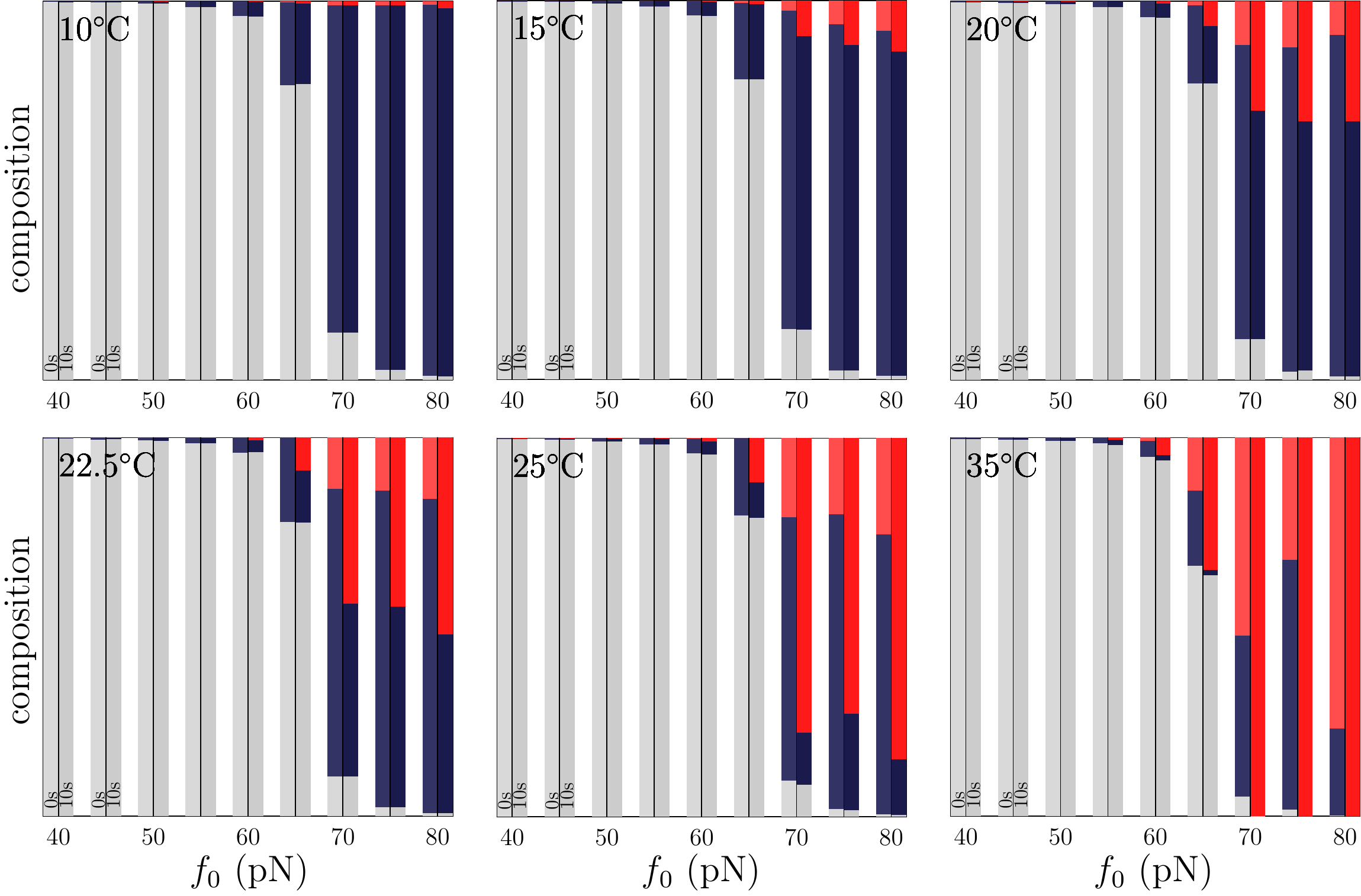} 
\caption{\label{fig5} (Color) `Dynamic phase diagram' for 40 kbp fragments of $\lambda$-DNA with one free end, at salt concentration $c=$500 mM. The dynamic stretching protocol is as Fig.~\ref{fig3}: note the larger fraction of S-DNA seen here.}
\end{figure}

\begin{figure}
\label{}
\centering
\includegraphics[width=0.9\linewidth]{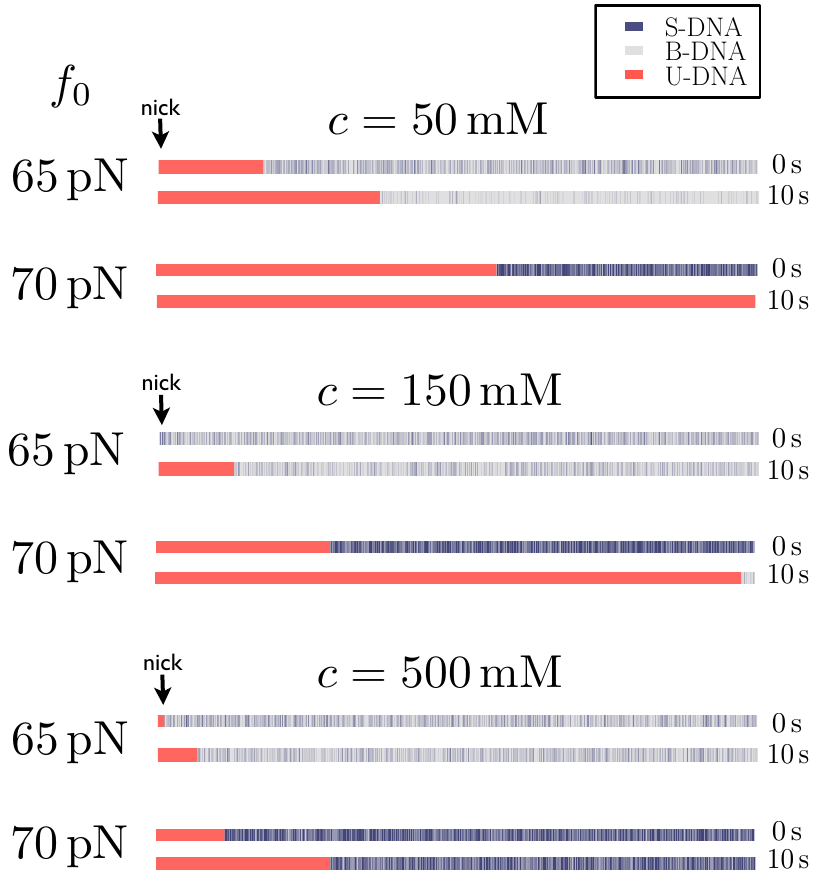} 
\caption{\label{fig6} (Color) Molecular compositions at 20$^{\circ}$C as a function of sequence (horizontal axis) for 40 kbp fragments of $\lambda$-DNA stretched to two target forces $f_0$ at three salt concentrations $c$. The stretching protocol is as Fig.~\ref{fig3} (force and molecular extension may vary after zeroing the loading rate). Each molecule possesses one free end. Low salt conditions and long waiting times favor unpeeling; at high salt conditions and short waiting times one is more likely to observe S-DNA.}
\end{figure}

\noindent
Direct sampling of unpeeling dynamics within $\lambda$-DNA is prohibitively slow near B/U phase coexistence, because the junction separating unhybridized and hybridized regions of the molecule is akin to a random walker in a free energy landscape replete with barriers in excess of hundreds of $k_{\rm B}T$~\cite{cocco}. We therefore computed thermal averages in Fig.~\ref{fig1}(b) within a constrained ensemble (at constant force and temperature) in which the molecule possessed one free end, permitting unpeeling, and melting was disallowed. Thermal averages were computed from $\langle A \rangle \equiv   \sum_n g_n(f,T) A(n) e^{-\beta F(n)}/\sum_n g_n(f,T) e^{-\beta F(n)}$, with $F(n) \equiv (L-n) \langle F \rangle_{\rm hyb.}+n F_{\rm  U}+\epsilon_{{\rm BU}} \delta_{n,0}$. Here $n\in[0,L]$ is the number of unpeeled basepairs; $F_{\rm  U}$ is the bulk internal energy per basepair of state U (see Equation (5) of Ref.~\cite{us}); $\epsilon_{\rm BU}<0$ is the U/B surface tension parameter; and $\langle F \rangle_{\rm hyb.}$ is the bulk energy per basepair of a molecule that is constrained to remain hybridized (calculated by direct Monte Carlo sampling). The degeneracy $g_n(f,T)$ accounts for the mixing entropy of B- and S-DNA within this hybridized phase. Here we have used the approximation $g_n(f,T) = \exp \left\{ (1-n/L) \ln \binom{L}{m} \right\}$, where $m$ is half the thermal average of the number of S/B domain walls.

Sequence averaging in Fig.~\ref{fig1} was done by dividing the basepairing-stacking parameters of the melting model of Ref.~\cite{santa} into two sets, containing 1) any A or T bases, or 2) any C or G bases. As in Ref.~\cite{us}, this leads to bulk free energies (at $f=0$) of melting of $0.67+0.19\, \ln(M/0.150)$ (AT) and $2.78+0.19\, \ln(M/0.150)$ (CG), in units of $\kt$ at 310 K. Here $M$ is the molar NaCl concentration (we use the salt-dependent correction from~\cite{cocco}).

\clearpage

\linespread{1}

\break

\break

\break

\break

\end{document}